\documentclass[10pt,twocolumn]{revtex4}

\usepackage{graphicx}
\usepackage{amsmath}
\usepackage{mathrsfs,amssymb}
\usepackage{color}
\usepackage{multirow}
\usepackage{epstopdf}

\begin{document}

\title{All-optical measurement of high-order fractional molecular echoes by high-order harmonic generation}

\author{Baoning Wang,$^1$ Lixin He,$^{1,*}$ Yanqing He,$^1$ Yinfu Zhang,$^1$  Renzhi Shao,$^1$ Pengfei Lan,$^{1,*}$ and Peixiang Lu$^{1,2}$}

\address{{1}Wuhan National Laboratory for Optoelectronics and School of Physics, Huazhong University of Science and Technology, Wuhan 430074, China\\
{2}Hubei Key Laboratory of Optical Information and Pattern Recognition, Wuhan Institute of Technology, Wuhan 430205, China}

\email{*Corresponding author: helx_ hust@mail.hust.edu.cn,pengfeilan@mail.hust.edu.cn} 



\begin{abstract}
An all-optical measurement of high-order fractional molecular echoes is demonstrated by using high-order harmonic generation (HHG). Excited by a pair of time-delayed short laser pulses, the signatures of full and high order fractional (1/2 and 1/3) alignment echoes are observed in the HHG signals measured from CO$_2$ molecules at various time delays of the probe pulse. By increasing the time delay of the pump pulses, much higher order fractional (1/4) alignment echo is also observed in N$_2$O molecules. With an analytic model based on the  impulsive approximation, the spatiotemporal dynamics of the echo process are retrieved from the experiment. Compared to the typical molecular alignment revivals, high-order fractional molecular echoes are demonstrated to dephase more rapidly, which will open a new route towards the ultrashort-time measurement. The proposed HHG method paves an efficient way for accessing the high-order fractional echoes in molecules.
\end{abstract}
\maketitle

\section{Introduction}


Echo is a common phenomenon that occurs in a  nonlinear system  excited by a pair of delayed  perturbation. Since Hahn reported the first observation of echo response in a nuclear spin system in 1950 \cite{echo_area_0}, echo phenomenon has been discovered in many other nonlinear systems, including cyclotron beams \cite{echo_area_1,echo_area_2}, plasma waves \cite{echo_area_3}, photon \cite{echo_area_4}, cold atoms \cite{echo_area_5,echo_area_6}, cavity QED \cite{echo_area_7}, and hadron colliders \cite{echo_area_8,echo_area_9}. The echo effects are demonstrated to have significant applications in  magnetic resonance imaging (MRI) \cite{eei1}, 2D electronic \cite{eei2,eei3,eei4}, vibrational \cite{eei5,eei6,eei7}, and rotational spectroscopy \cite{align_echo_1,align_echo_2,align_echo_3,ddm,esd1}, and also the generation of short-wavelength radiation in free-electron lasers \cite{EEHG1,EEHG2,EEHG3}. 

Recently, a new type of echo, i.e., the molecular alignment echo, was introduced. Excited by two time-delayed short laser pulses, the echo response was first experimentally observed in CO$_2$ gas by measuring the laser-induced birefringence signal \cite{align_echo_1}. Later on, molecular echoes have also been observed in CH$_3$I \cite{align_echo_4} and OCS \cite{align_echo_5} molecules. Unlike the conventional molecular rotational revivals, the alignment echoes appear at the time delays of $\tau=NT$, where $T$ is the time delay between the two pump pulses, $N$ is a positive integer. The formation of these echoes can be theoretically explained as a consequence of the kick-induced filamentation of the rotational phase space \cite{align_echo_1,align_echo_2,align_echo_3}. In addition to the echoes at $\tau=NT$ (also called the full echoes), G. Karras \textit{et al}. also predicted some additional remarkable recurrences (which are referred to fractional echoes) at times that are rational fractions of the delay between the pump pulses \cite{align_echo_1,align_echo_2}. To observe the fractional echoes, it requires measurement of higher order moments of the molecular angular distribution, which however cannot be achieved in the birefringence measurement. Very recently, the lowest-order (1/2) fractional echo in molecules was optically detected by measuring the third harmonic of the probe pulse \cite{align_echo_2}. While Lin \textit{et al.} have also demonstrated the measurement of fractional molecular echoes (up to 1/3) with Coulomb explosion imaging (CEI) method \cite{align_echo_3}, a purely optical technique to probe high-order fractional molecular echoes is still lacking.

High-order harmonic generation (HHG) is a highly non-linear phenomenon that occurs in laser-driven recollision process \cite{3s1,3s2}. The generated harmonic spectrum encodes abundant information of structure and dynamics of the target, which has been widely employed for the ultrafast detection in atom \cite{at1,at1_2,at3_2,at4_3,at4_4,at4_5}, molecules \cite{mo1,HOMO2,mo3,mo3_2,mo4,mo5,mo7,mo8,mo8_2,mo8_3} and solid \cite{so1,so2,so3,so3_2,so3_3,so4,so4_2,so4_2_0,so4_2_1,so4_2_2,so4_2_3,so4_3,so4_5}. In particular, HHG from the molecular ensembles intrinsically convolutes the molecular angular distribution \cite{mad1,mad2,mad3,mad3_2,mad4}, which therefore is expected to be a powerful candidate for the detection of high-order molecular echoes.

In this work, we demonstrate the measurement of high-order fractional molecular echoes through HHG. Excited by a pair of time-delayed laser pulses, the full, 1/2, and 1/3 alignment echoes are observed in measured time-dependent HHG signals from CO$_2$ molecules. Higher-order fractional (reaching 1/4) echo is also observed in HHG from N$_2$O by increasing the time delay of the two pump pulses. The rotational dynamics of the echo response is further successfully retrieved from experimental results by using an analytic model based on the impulsive approximation.


\section{Results and discussion}

\begin{figure}[t]
	\includegraphics[width=\linewidth]{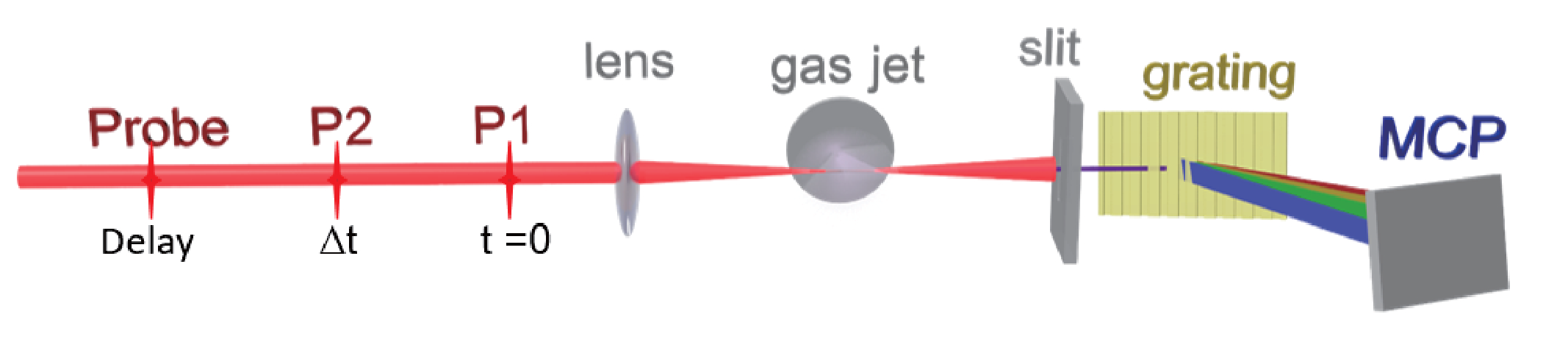}
	\caption{\label{fig1} Experiment setup of harmonic generation from molecular ensemble excited by a pair of time-delayed laser pulses.
	}
	\label{fig:moving vs static}
\end{figure}

The experiment is conducted by using a Ti:sapphire laser system (Astrella-USP-1K, Coherent, Inc.), which delivers 35 fs, 800 nm laser pulses at a repetition rate of 1 kHz. The experiment setup is sketched in Fig. \ref{fig1}. The output laser pulses are divided into three optical paths to generate two pump pulses (P$_1$ and P$_2$) and one probe pulse. The three laser pulses are linear polarized and parallel to each other in our experiment. As shown in Fig. \ref{fig1}, the first pump laser pulse P$_1$ is applied at $t=0$ (the initial time throughout the paper), which leads to a transient alignment of molecular ensemble within a narrow cone. Afterwards the transient alignment vanishes due to the dispersion of molecular angular velocities. The second pump laser pulse (P$_2$) is applied at a delayed time $\Delta t$, resulting in another transient excitation. As reported in  \cite{align_echo_1,align_echo_2,align_echo_3,align_echo_4,align_echo_5}, excited by these two pump pulses, the alignment echoes will be formed at certain time delays after P$_2$. To probe the following echo dynamics, an intense laser pulse is applied to interact with the excited molecules to generate high-order harmonics at various time delays (relative to P$_1$). These three incident laser pulses are focused into a 500$\mu m$ continuous gas valve by a lens (f=400 mm. The background gas pressure is 20 Torr and the rotational temperature at the focus is estimated to be 100K in terms of the procedure introduced in \cite{rt1,rt2}. The generated high harmonic spectrum is recorded by a home-made spectrometer consisting of a flat-field grating (1200 grave/mm), a multichannel-plate detector backed with a phosphor screen, and a charge coupled device (CCD) camera \cite{mad3}.

\begin{figure*}[t]
	\centering
	\includegraphics[width=.8\linewidth]{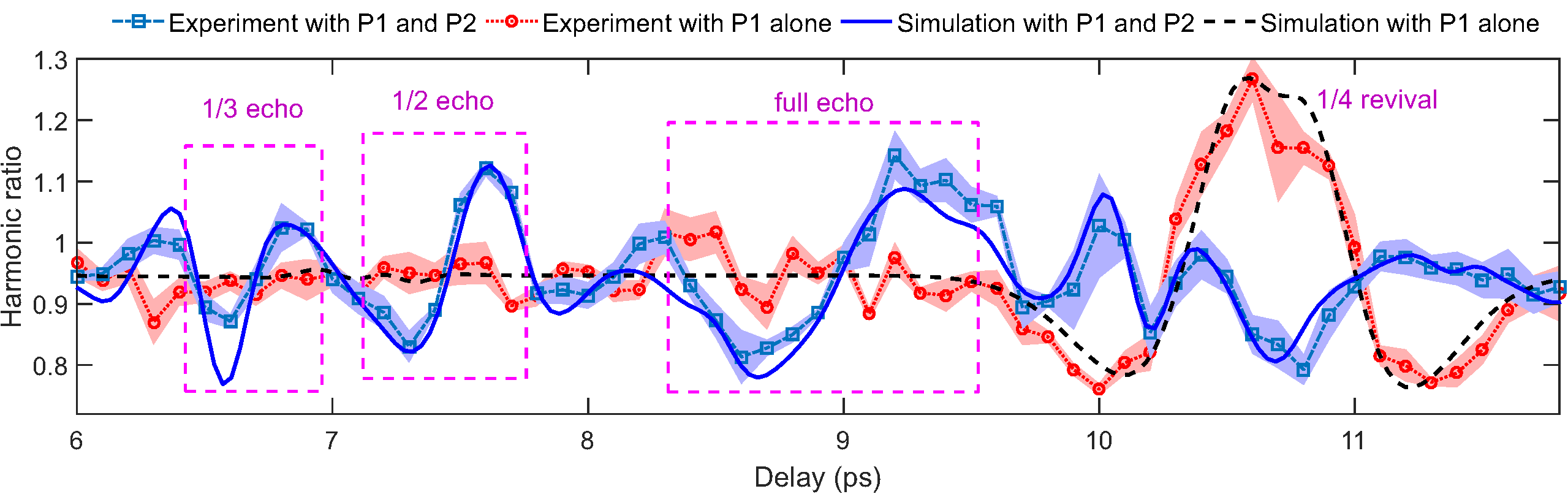}
	\caption{\label{fig2} Measured time-dependent harmonic signals from the CO$_2$ molecules excited by pump P$_1$ and P$_2$ (blue squares with dashed line). For comparison, the results from the molecules excited by P$_1$ alone are also presented (red circles with dotted line). The dashed rectangles show the locations of the full, 1/2, and 1/3 echoes. Shaded areas represent the standard deviation of five independent measurements. Blue solid and black dashed lines are the simulated results with the optimized laser intensities of P$_1$ and P$_2$.
	}
	\label{fig:mechanism}
\end{figure*}


We first perform the experiment with CO$_2$ gases. In our experiment, the time delay between P$_1$ and P$_2$ is set to $\Delta t$=4.8 ps, and the laser energies of pump pulses P$_1$, P$_2$ and probe pulse are 143 $\mu$J, 52 $\mu$J and 560 $\mu$J, respectively. Figure \ref{fig2} presents the time-dependent harmonic signals of 25th harmonic (H25) generated from the excited molecules (the blue squares with dashed line). Here, the abscissa denotes the time delay $t$ between the probe pulse and the first pump pulse P$_1$, and the harmonic signals have been normalized by the results from the random molecules (i.e., no pump pulses). For comparison, the result from the molecules pumped by P$_1$ pulse alone is also presented as red circles with dashed line in Fig. \ref{fig2}. Excited by P$_1$ alone (i.e., a typical molecular alignment situation), the alignment events mainly occur at quarter rotational revivals, i.e., $\frac{1}{4}$T$_{rev}$, $\frac{1}{2}$T$_{rev}$, etc. (T$_{rev}\sim42.7 $ps is the revival period of CO$_2$). From Fig. \ref{fig2}, one can clearly see the alignment signal of the first quarter quantum revival at the time delay of 10.7 ps.
While within the interval of two adjacent quarter revivals (like the delays of 6 ps-9.7 ps, which is between 0 and $\frac{1}{4}$T$_{rev}$), the molecular ensemble is nearly isotropic, therefore the harmonic signals exhibit no modulation. When P$_2$ is applied at $\Delta t=$4.8 ps, one can see significant modulations in the HHG signals at the time delays of 8.4 ps-9.6 ps, 7 ps-7.8 ps, and 6.5 ps-7 ps (indicated by dashed rectangles in Fig. \ref{fig2}), which just correspond to the appearance time of the full, 1/2, and 1/3 echoes, respectively. Similar results have also been observed in other harmonics and even a spectrally integrated signal. These results mean that the echo response has been recorded in the time-dependent HHG signals. Besides, compared to the typical alignment excited by P$_1$ alone, the HHG signal excited by both P$_1$ and P$_2$ is severely modulated and shows a more rapid oscillation at the quarter revival (10 ps-11.2 ps), indicating a significant influence of the pump pulse P$_2$ on the molecular rotational dynamics.

Next, we demonstrate to extract the molecular rotational dynamics from the measured HHG signals. HHG from the rotational molecules is a convolution of the molecular angular distribution and the electronic response on the single-molecule level. Since the HHG process is much faster than the molecular rotation, molecular rotation during the probe pulse is usually neglected \cite{QRS1,QRS2}. The time-dependent harmonic signals from the rotational molecules can be given by coherent superposition of emission from the molecules aligned at different angles \cite{mad3,QRS1}, 
\begin{align}
S(\alpha,t) = |\int_{\phi=0}^{2\pi}\int_{\theta=0}^{\pi}D_q(\theta)\rho(\theta,\phi,t)sin\theta d\theta d\phi|^2.
\label{HHGsim}
\end{align}
Here,  $D_q(\theta)$ is the angle-dependent induced dipole moment of $q$th harmonic on the single-molecule level. $\rho(\theta,\phi,t)$ is the transient molecular angular distribution, which is related to spatiotemporal evolution of molecular rotational wave packet. $\theta$ and $\phi$ are the polar and azimuthal angles of the molecular axis. Unlike the CEI approach that measures the time-dependent molecular distributions directly, the extraction of the molecular distributions from HHG signals requires a prior knowledge of the laser-induced dipole moment. 
To ensure the precision of the extraction, the single-molecule HHG process should be  accurately described. 
In this work, we use the quantitative rescattering (QRS) theory \cite{QRS1,QRS2} to model the single-molecule HHG process.  The accuracy and validity of QRS have been well established by comparing with the ``exact'' numerical solution of the time-dependent
Schr\"{o}dinger equation (TDSE) and the experimental results \cite{QRSV1,QRSV2,QRSV3}. 
Within QRS theory, the laser-induced dipole moment for fixed-in-space molecule is given by $D_q(\theta)=N(\theta)^{\frac{1}{2}}W(q,\theta)d(q,\theta)$. Here, $N(\theta)$ is the alignment-dependent ionization rate, which is calculated by the molecular Ammosov-Delone-Krainov (MO-ADK) theory \cite{MOADK}. $W(q,\theta)$ is called the recolliding electron wave packet, which is obtained by solving the TDSE with a reference atom that has nearly the same ionization potential as the molecule \cite{QRS1}. The photorecombination transition dipole $d(q,\theta)$ is given by $d(\omega)=\langle \Psi_0|r|\Psi_f\rangle$, where $\Psi_0$ is the initial bound state and $\Psi_f$ is the final continuum state. In our simulation, $\Psi_0$ is obtained from the MOLPRO code \cite{QRSd1} within the valence complete-active-space self-consistent field method. The final state $\Psi_f$ is described in a single-channel approximation, which is given by $\Psi_f=A[\varPhi\phi_k]$. Here, $\varPhi$ is the correlated electron wavefunction of the parent ion, $\phi_k$ is the continuum electron wavefunction, and $A$ is the antisymmetrization operator. The target part $\varPhi$ is calculated by a valence complete-active-space configuration interaction wavefunction using the same bound states in the initial state. The continuum wavefunction $\phi_k$ is calculated by solving the Schr\"{o}dinger equation for the continuum electron with the iterative Schwinger variational method \cite{QRSd2}. 
In our simulation, the intensity of the probe pulse is estimated from the measured harmonic cutoff (the 31st harmonic), which is about 1.85$\times$10$^{14}$W/cm$^2$. 
According to Eq. \eqref{HHGsim}, the HHG signals depend on the angle dependence of the single-molecular dipole. We can deduce that a monotone angle-dependent single-molecule dipole (both amplitude and phase) with a large difference between 0$^\circ$  and 90$^\circ$  alignment would be more beneficial to the observation of the echo signals. In contrast, a flat angle dependence of the single-molecule dipole will be detrimental to the observation of the echoes.

Considering the durations of the pump pulses are much shorter than the typical timescale of molecular rotation, we adopt the impulsive approximation and treat the pump pulses as $\delta$-pulse \cite{dp1}. Based on this assumption, the rotation wave function resulting from the interaction between pump P$_1$ and the initial molecular rotational state $\Psi_{JM}^0$ (the eigenstate $|JM\rangle $ of field-free rigid rotor described by the spherical harmonic $Y_J^M(\theta,\phi)$) can be written as,
\begin{align}
\Psi'_{JM}=e^{i\mathscr{P}_1cos^2\theta}\Psi_{JM}^0.
\end{align}
Here, $\mathscr{P}_1=\frac{1}{4}\Delta\alpha\int E_1(t)^2dt$ is the dimensionless strength of the first pump laser pulse P$_1$, $E_1(t)$ is the laser envelop of the pump pulse P$_1$. The propagation of the generated molecular rotational wave packet in time is then calculated by the spectral decomposition of the wave function \cite{dp1}:
\begin{align}
\Psi'_{JM}(\theta,\phi,t)=\mathop{\Sigma}_{J'}C'_{JM,J'M}e^{-i\frac{E_{J'}}{\hbar}t}Y_{J'}^{M}(\theta,\phi).
\end{align}
Here, $E_{J'}$ is the eigenenergy of the eigenstate $|J'M\rangle $. Note that, excited by the linearly polarized pump laser, the magnetic quantum number M is conserved during the rotational transition. When P$_2$ is applied at the delay $\Delta t$, a new rotational wave function $\Psi''_{JM}$ is generated,
\begin{align}
\Psi''_{JM}=e^{i\mathscr{P}_2cos^2\theta}\Psi'_{JM}(\theta,\phi,\Delta t).
\end{align}
Here, $\mathscr{P}_2=\frac{1}{4}\Delta\alpha\int E_2(t)^2dt$ is the dimensionless strength of the second pump laser pulse P$_2$, $E_2(t)$ is the laser envelop of the pump laser P$_2$. Similarly, the propagation of the new generated molecular rotational wave packet in time can be calculated as
\begin{align}
\Psi''_{JM}(\theta,\phi,t)=\mathop{\Sigma}_{J''M}C''_{JM,J''M}e^{-i\frac{E_{J''}}{\hbar}(t-\Delta t)}Y_{J''}^{M}(\theta,\phi).
\end{align}
The molecular angular distribution $\rho(\theta,\phi,t)$ is given by the weighted average of the modulus squares of the molecular rotational wave packet:
\begin{align}
\rho(\theta,\phi,t)=\mathop{\Sigma}_{JM}\Gamma_{JM}|\Psi''_{JM}(\theta,\phi,t)|^2.
\end{align}
Here, $\Gamma_{JM}$ is the statistic weight of initial rotational state $|JM\rangle $ according to the Boltzmann distribution.

\begin{figure}[t]
	\centering
	\includegraphics[width=\linewidth]{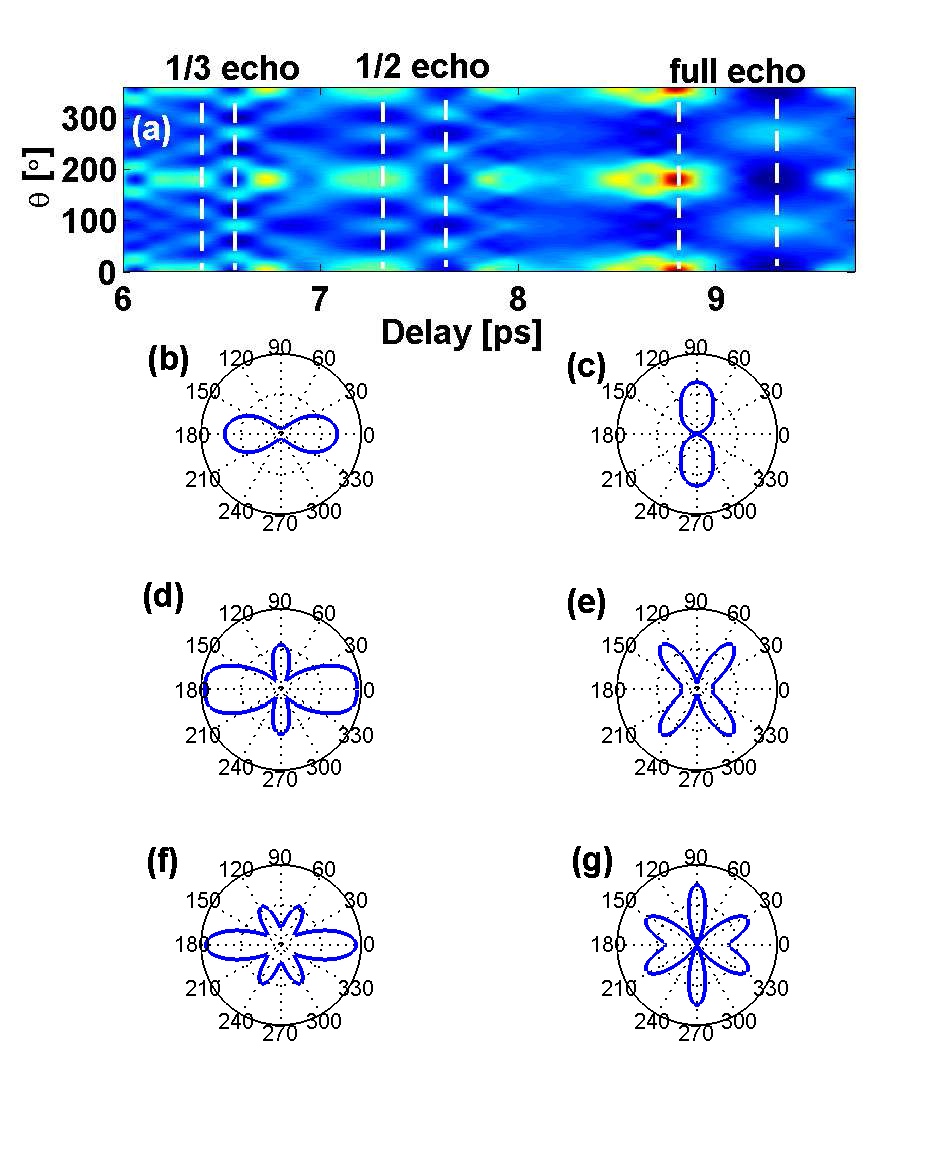}
	\caption{\label{fig3} (a) The time-dependent angular distribution of molecular rotational wavepacket of CO$_2$ molecule simulated with the optimal intensities of P$_1$ and P$_2$ (I$_1$=4$\times10^{13}$W/cm$^2$ and I$_2$=1.5$\times10^{13}$W/cm$^2$). (b)-(c) Polar plots of the angular distributions of full echo at t=8.7 ps (in the left column) and t=9.2 ps (in the right column). (d)-(e) Same as (b)-(c), but for 1/2 echo at t=7.3 ps and t=7.6 ps. (f)-(g) Same as (b)-(c), but for 1/3 echo at t=6.6 ps and t=6.8 ps. 
	}
	\label{fig:scalingLaw_CREI}
\end{figure}


\begin{figure*}[t]
	\centering
	\includegraphics[width=.8\linewidth]{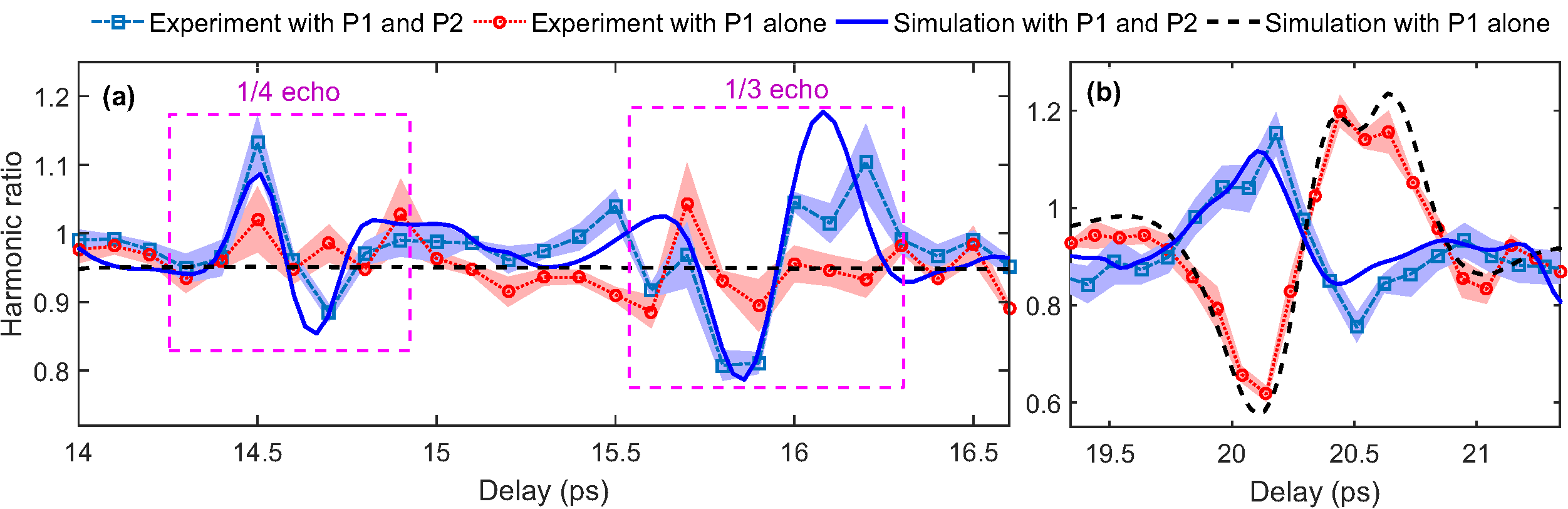}
	\caption{\label{fig5} (a) Measured time-dependent harmonic signals (H25) from the N$_2$O molecules excited by pump P$_1$ and P$_2$ (blue squares with dashed line). For comparison, the results from the molecules excited by P$_1$ alone are also presented (red circles with dotted line). The dashed rectangles show the locations of the 1/3 and 1/4 echoes. Shaded areas represent the standard deviation of five independent measurements. Blue solid and black dashed lines are the simulated results with the optimized laser intensities of P$_1$ and P$_2$. (b) Same as (a), but for the results measured at the half quantum revival. 
	}
	\label{fig:compareMass}
\end{figure*}


In our experiment, the intensities of the two pump pulses are not certain. Equations (1)-(6) build connection between the intensities of these two pump pulses and the harmonic signals. To retrieve the molecular rotational dynamics, we have performed simulations with different intensity combinations of the two pump pulses according to Eqs. (1)-(6). Note that, in order to produced strong echo signals, in our experiment and therefore the simulations, the intensity of P$_2$ is maintained smaller than P$_1$ \cite{align_echo_1,align_echo_2,align_echo_5}. The optimal intensity combination of P$_1$ and P$_2$ is I$_1$=4$\times10^{13}$W/cm$^2$ and I$_2$=1.5$\times10^{13}$W/cm$^2$, which is obtained by minimizing the squared difference between the simulated and measured harmonic signals. Using the optimal intensities of P$_1$ and P$_2$, we have calculated the time-dependent harmonic signals as presented in Fig. \ref{fig2} (blue solid and black dashed lines), which agree well with the experimental results. Here, it's worth mentioning that in our simulation only the highest occupied molecular orbital (HOMO) of CO$_2$ is considered. It has been reported that the multi-orbital contribution could play an important role in HHG from CO$_2$ \cite{HOMO2}. However, in our work (and also previous works \cite{QRS1,QRS2}), the HOMO of CO$_2$ alone can well reproduce the experimental results. The discrepancy between the two interpretations is likely due to the accuracy of photoionization cross section (PICS) used in the theory. In QRS, accurate transition dipoles from state-of-the-art molecular photoionization calculation were used \cite{QRS1,QRS2}. The PICS from these calculations have been found to be in good agreement with conventional photoionization experiment.

We further investigate the echo dynamics by analyzing the spatiotemporal evolution of molecular rotational wave packet. Figure \ref{fig3} shows the time-dependent molecular angular distribution simulated with the optimal intensities of P$_1$ and P$_2$. One can see clear signatures of the full, 1/2, and 1/3 echoes at 8.7-9.2 ps, 7.3-7.6 ps, and 6.6-6.8 ps [indicated by the dashed lines in Fig. \ref{fig3}(a)], respectively. More clearly, we present the polar plots of the angular distributions of these echoes in Figs. \ref{fig3}(b)-\ref{fig3}(g). As shown in Figs. \ref{fig3}(b)-\ref{fig3}(c), the full echo exhibits a cigar-shaped angular distribution and transforms from alignment at t=8.7 ps [Fig. \ref{fig3}(b)] to anti-alignment at t=9.2 ps [Fig. \ref{fig3}(c)]. The 1/2 echo is demonstrated to have a cross-shaped distribution at t=7.3 ps [Fig. \ref{fig3}(d)] and then a butterfly structure at t=7.6 ps [Fig. \ref{fig3}(e)]. For 1/3 echo, it shows much more complex six-lobe structures at the alignment [t=6.6 ps in Fig. \ref{fig3}(f)] and anti-alignment [t=6.8 ps in Fig. \ref{fig3}(g)] moments. All these results are consistent with previous studies \cite{align_echo_3}. Besides, from Fig. \ref{fig3}(a), one can see that higher-order fractional molecular echoes experience a much rapid transition from alignment to anti-alignment. Such a rapid transition will be beneficial for the ultrashort-time measurement in molecules \cite{align_echo_2,ddm}.

As shown in Fig. \ref{fig2}, the time space between the 1/2 and 1/3 echo is very narrow due to the small time delay between the pump pulses. In order to visualize higher-order fractional molecular echoes, one can increase the delay of the pump pulses. However, for CO$_2$, the interplay of quantum revival and echo phenomenon will show up when the delay of the pump pulses is further increased \cite{align_echo_2,align_echo_3}. Alternatively, we adopt the N$_2$O as the sample gas, which has a similar revival period (T$_{rev}$=39.9 ps) as CO$_2$. However, unlike CO$_2$, N$_2$O does not demonstrate quarter revivals due to its different symmetry, which therefore provides much longer time window for the observation of the higher-order fractional molecular echoes. Figure \ref{fig5}(a) presents the harmonic signals (blue squares with dashed line) measured from N$_2$O with the time delay between P$_1$ and P$_2$ set to $\Delta t=$12 ps. The results pumped by P$_1$ alone are also presented for comparison [red circles with dashed line in Fig. \ref{fig5}(a)]. One can see clear signature of 1/4 echo in the range from 14.3 ps to 15 ps [indicated by the dashed rectangle in Fig. \ref{fig5}(a)]. Similar results have also been observed in other harmonic orders, e.g., H23 and H27, etc. Moreover, we have also performed above-mentioned simulation process for N$_2$O and the results are presented in Fig. \ref{fig5}(a) (blue solid and black dashed lines), which are in good agreement with the experiment. Besides, 
the HHG signals at the half revival of N$_2$O have also been measured. As shown in Fig. \ref{fig5}(b), excited by both P$_1$ and P$_2$, the time-dependent HHG signals (blue squares with dashed line) present a reversed modulation compared to that by P$_1$ alone (red circles with dotted line).



\section{Conclusion}


In summary, we have realized an all-optical measurement of higher-order fractional molecular echoes by using HHG. The full, 1/2, and 1/3 echoes are observed by measuring the time-dependent HHG from CO$_2$ gases excited by a pair of time-delayed laser pulses. Using an analytic model based on the impulsive approximation, the spatiotemporal dynamics of the echo response are extracted, which agree well with the recent CEI measurement \cite{align_echo_3}. Moreover, the 1/4 echo is observed in HHG from N$_2$O molecules with a longer time delay of the pump pulses. Higher-order fractional molecular echoes are demonstrated to transform more rapidly, which can be extended to the ultrashort-time measurement of the dissipative and decoherence processes in molecules \cite{ddm,dp1_2} by using a mid-IR laser \cite{dp3} with high gas pressure \cite{dp2_2}.


\section*{Funding}
National Natural Science Foundation of China (No. 11704137, 11627809, 11874165, 11774109); National key research and development program (2017YFE0116600); Fundamental Research Funds for the Central Universities (2017KFXKJC002); Science and
Technology Planning Project of Guangdong Province (2018B090944001); the Program for HUST Academic Frontier Youth Team.

\section*{Disclosures}
The authors declare no conflicts of interest.






\end{document}